\documentclass[10pt, a4paper]{article}

\usepackage{endnotes}
\let\footnote=\endnote

\usepackage[utf8x]{inputenc}
\usepackage{graphicx}
\usepackage{authblk}

\date{}

\title{Quantifying time-dependent Media Agenda and Public Opinion by topic modeling}

\author[1,2]{Sebasti\'an Pinto \thanks{spinto@df.uba.ar}}
\author[3]{Federico Albanese}
\author[1,2]{Claudio O. Dorso}
\author[1,2]{Pablo Balenzuela}

\affil[1]{\small{Departamento de F\'isica, Facultad de Ciencias Exactas y Naturales, Universidad de Buenos Aires, Av.Cantilo s/n, Pabell\'on 1, Ciudad Universitaria, 1428, Buenos Aires, Argentina.}}
\affil[2]{Instituto de F\'isica de Buenos Aires (IFIBA), CONICET, Av.Cantilo s/n, Pabell\'on 1, Ciudad Universitaria, 1428, Buenos Aires, Argentina.}
\affil[3]{Instituto de Investigaci\'on en Ciencias de la Computaci\'on (ICC), CONICET, Av.Cantilo s/n, Pabell\'on 1, Ciudad Universitaria, 1428, Buenos Aires, Argentina.}

\begin{document}

\maketitle

\begin{abstract}
The mass media plays a fundamental role in the formation of public opinion, either by defining the topics of discussion or by making an emphasis on certain issues. Directly or indirectly, people get informed by consuming news from the media. Naturally, two questions appear: What are the dynamics of the agenda and how the people become interested in their different topics?
These questions cannot be answered without proper quantitative measures of agenda dynamics and public attention. 
In this work we study the agenda of newspapers in comparison with public interests by performing topic detection over the news. We define Media Agenda as the distribution of topic's coverage by the newspapers and Public Agenda as the distribution of public interest in the same topic space.
We measure agenda diversity as a function of time using the Shannon entropy and differences between agendas using the Jensen-Shannon distance. We found that the Public Agenda is less diverse than the Media Agenda, especially when there is a very attractive topic and the audience naturally focuses only on this one. Using the same methodology we detect coverage bias in newspapers.
Finally, it was possible to identify a complex agenda-setting dynamics within a given topic where the least sold newspaper triggered a public debate via a positive feedback mechanism with social networks discussions which install the issue in the Media Agenda.
\end{abstract}

\section{Introduction}

\par One of the challenges in complex social system research is to understand the ecosystem of information flow and opinion formation. A major role within this ecosystem is played by the mass media outlets, which are massively used as sources of information.
People get informed by the media and then interact among them via personal discussions or through social networks, giving rise to a complex dynamics where opinions are shaping and changing with time. In this scenario, it becomes essential to understand the influence of mass media in a given social group.
\par The influence of the media on public opinion was first explored by the social sciences. In the seminal study performed in \cite{mccombs1972agenda}, Maxwell McCombs and Donald Shaw found that the aspects of public affairs that are prominent in the news become prominent in the public. This work is considered the founding of the agenda-setting theory.
In its basic stage, known as first-level agenda-setting \cite{mccombs2005look}, the theory focuses on the comparison between the topics coverage by the media (Media Agenda) and those that the public consider as priority (Public Agenda).
For instance, within the agenda-setting framework, it was explored how media content correlates with audiences of different ages \cite{coleman2007young} and how people agendas differ based on the way they consume news \cite{althaus2002agenda}.
On the other hand, the theory hypothesizes how the media affects the audience opinion, in particular, how political coverage and political advertisement shape candidate knowledge among the audience \cite{brians1996campaign, gerber2009does}, or how the coverage given by the media to a particular nation affects people perception about its importance to local political interests \cite{wanta2004agenda}.
Other works examine the differences between public and journalists preferences \cite{mitchelstein2016brecha}, or study the coverage of the main newspapers on particular events related to a confrontation scenario between government and press \cite{zunino2010cobertura, koziner2013cobertura}.
\par The agenda-setting theory also induced other several research directions \cite{mccombs2005look}. One of them focuses on detecting media bias, either by taking into account the number of mentions related to a preferred political party \cite{lazaridou2016identifying, baumgartner2015all} or by identifying the ideology through the position of the media regards to certain issues or actors \cite{elejalde2018nature, sagarzazu2017hugo}. 
Since the irruption of internet, a quantitative analysis based on the access to big data has become available, as for instance, those who take into account temporal dependence of the media and public attention. 
In \cite{soroka2017negativity}, it is shown that the newspapers and Twitter have an opposite reaction to the changes of the unemployment rates; in \cite{guggenheim2015dynamics}, the competition of frames about gun control is explored; in \cite{ali2018measuring}, the authors show how fluctuations of Twitter activity in different regions depend on the location of terrorist attacks; and in \cite{russell2014dynamics}, the complex interplay between the social media and the traditional one is followed over time on a set of predefined, but general issues.
\par It is important to notice that the articles cited above work either on a single issue or on a set of predefined issues usually selected by the researcher. However, a data driven selection of issues can be performed using a tool frequently employed in the analysis of large document corpus: Unsupervised topic modeling. 
It is an alternative to the dictionary-based analysis, which is the most popular automated analysis approach \cite{guo2016big}, and allows to work with a corpus without a prior knowledge, letting the topics emerge from the data. 
In the same spirit of our work, many authors, as for instance \cite{mohr2013introduction, gunther2016word, jacobi2016quantitative}, emphasized the advantage of using automated text classification in social science research.
For example, topic modeling have been used in discourse analysis \cite{tornberg2016muslims}, analysis of social media discussions \cite{paul2014discovering, malik2016macroscopic, kim2016topic}, or recognition of entities in news articles \cite{newman2006analyzing}.
However, much works on topic models are also devoted to the performance of the topic model over a labeled corpus, focusing on the proper detection of the topics \cite{dai2010online, po2016topic, brun2000experiment}, and in general, issues about the temporal profile of topics are embedded in the context of topic tracking \cite{hu2016news, li2017joint}, or in the recognition of emerging topics in real-time \cite{cataldi2010emerging}, mostly applied to social media. 
\par In this work, we propose to perform an unsupervised topic model on newspapers articles to study the dynamics of Mass Media and Public Agendas. We define Media Agenda as time-evolving distribution of topic's coverage by different media outlets and Public Agenda as the distribution of public interest in the same topic space, by looking both the Google searches and Twitter activity. We apply this method to study the dynamics of the Argentinian media agenda due to our familiarity with the political background, but as can be seen throughout the work, the methodology implemented is far general and can be easily extended to other datasets. 

\par This work is organized as follows: In section \ref{sec:MatMeth} we describe the analyzed corpus of news articles, their numerical representation, the used topic model and the definitions of media and public agendas. In section \ref{sec:Results} we show the time-evolving agendas (subsection \ref{sec:Quantification}), and calculate their diversity (subsection \ref{sec:Diversity}) and the distance among them (subsection \ref{sec:Distance}). In subsection \ref{sec:Agenda-bias} we study the differences among newspapers agendas and in subsection \ref{sec:single-topic} we devote to a single-topic analysis. 
Finally, in section \ref{sec:Conclusions} we draw the conclusions and perspectives of this work.

\section{Materials and Methods}
 \label{sec:MatMeth}

\subsection{The Media Agenda}
\par We analyze a three-month period of the Argentinian media agenda composed by a corpus of news articles that were published between July 31st, 2017 and November 5th, 2017. 
The articles come from the political section of the online editions of the Argentinian newspapers \emph{Clar\'in}, \emph{La Naci\'on}, \emph{P\'agina12}, and the news portal, \emph{Infobae}. The first two lead the sale of printed editions in Buenos Aires city, but Clar\'in reaches roughly two times the readers of La Naci\'on, and ten times the readers of P\'agina 12 \cite{IVC}, who was chosen because of its left political orientation. On the other hand, Infobae has the most visited website, much more than Clar\'in and La Naci\'on \cite{AlexaAR}.
The corpus analyzed is made up by 11815 news articles: 2908 of Clar\'in, 3565 of La Naci\'on, 3324 of P\'agina 12, and 2018 of Infobae. Except P\'agina 12, all articles were taken from the section \emph{Pol\'itica} (Politics) of the respective news portals, while the articles which belong to P\'agina 12 were taken from the section \emph{El pa\'is} (The country).
\par The articles are described as numerical vectors through the \emph{term frequency - inverse document frequency (tf-idf)} representation \cite{xu2003document}. The dimension of the vectors ($t=445993$) is given by the total amount of words of the corpus after removing non-informative ones such as prepositions and conjunctions. Each term of the vector gives the frequency this word appears in the document (tf) weighted by a factor (idf) who measures its degree of specificity (i.e., if this word is frequently used only in this document or in the whole corpus). A more detailed description can be found in \ref{sec:agenda_construction}.

\par Once the document vectors are constructed, we put them together in a document-term matrix (\emph{M}), which has dimensions of number of documents in the corpus ($d$) by number of terms ($t$). The next step is to find how the news articles can be grouped by similarity in clusters called topics. A topic is defined as a group of similar articles which roughly talks about the same subject.

\par In order to detect the main topics in the corpus, we perform \emph{non-negative matrix factorization (NMF)} \cite{xu2003document, lee1999learning} on the document-term matrix (\emph{M}). NMF is an unsupervised topic model which factorizes the matrix \emph{M} into two matrices \emph{W} and \emph{H} with the property that all three matrices have no negative elements (see Eq.~\ref{eq:nmf}).
This non-negativity makes the resulting matrices easier to inspect, and very suitable for topic detection because it provides the representation of the articles in the topic space (matrix \emph{H}) where each column gives the degree of membership of each article to a given topic. Also, the matrix \emph{W} provides the combination of words which describe each topic.

\par After performing NMF, we define the Media Agenda as the distribution of the topic's weights and their evolution in time. The weight of topic $i$ ($W_i$) is defined as the combination of the amount of news articles of the topic (weighted by their degree of membership) times the length of the article. It can be defined in a daily basis (time-dependent distribution) or for the whole three-month period (average distribution).
For a single day $d$, it is sketched in the Eq.~\ref{eq:topic_weight}, where $l(j)$ is the number of words of the document $j$; $h_{ji}$ (element of matrix \emph{H}) is the degree of membership of document $j$ on topic $i$; $d_j$ is the date of document $j$; and $\delta$ is the Kronecker delta. 
Providing by the fact that each document vector can have all non-zero components, it is allowed that a document contributes to more than one topic weights.
In order to reduce noise, we apply a linear filter with a three day wide sliding window, and finally we normalize the temporal profiles. 

\begin{equation}
W_i(d) = \sum_j l(j) \cdot h_{ji} \cdot \delta_{d_j,d}
\label{eq:topic_weight}
\end{equation}

\subsection{The Google and Twitter Agendas}
 
\par Besides the construction of the Media Agenda, it is important to have some measure of the public interests and construct what we call the Public Agenda. 
To achieve this goal, we take Google and Twitter as proxies of the public interests by looking for the same topics in the same period of time. We take advantage of the topic keywords in order to make queries into the Google Trends (\emph{GT}) tool and into the advance search tool of Twitter (\emph{Tw}), and therefore we get the relative weight of searches and tweets in each respective platform. The distribution of searches in Google Trends and number of tweets in Twitter is what we define as the Public Agendas.

In Table \ref{table:gt_queries} we point out the keywords involved in the queries, and in \ref{sec:agenda_construction} we provide a more detailed description of the methodology employed in the construction of the Public Agendas. Briefly, we only mention that we directly obtained the relative amount of keywords searches in Google Trends, while in Twitter we inferred that information by taking a sample of tweets per day. During all the studied period, we took a total of 24360 tweets.

Note that the way we construct the Public Agenda is defined by the topics found in the Media Agenda and therefore it would not be possible to find topics in public interests which were not published in media outlets in the analyzed three month period. However, the same intrinsic limitation provided by this methodology allows us to define both agendas in the same topic space and therefore perform proper comparison measures, as we will show in section \ref{sec:Results}.

\section{Results}
\label{sec:Results}

\subsection{The topic decomposition}
\label{sec:topic-decomposition}

\par We initially perform a topic decomposition of the three-month period corpus of news reported above by focusing on the ten most important issues. 
The reason behind factorizing the corpus in ten topics was based on having a low dimensional representation of the corpus and a clear interpretation of the topics due to our prior knowledge of the political background. We found that this factorization allowed us to draw useful conclusions. However, more sophisticated methodologies to estimate the number of topics in a corpus can be taken into account in future researches.

The keywords which define the ten topics are represented in the word clouds of Fig~\ref{fig:topics_wordclouds}.
Given our interpretation of the keywords found in three of them, we joined these topics as being part of the same macro-topic which we called \emph{Elections}.
On the other hand, the same holds for other two topics which were classified as part of a macro-topic called \emph{Missing person}. 
Therefore, the ten original topics were reduced to seven, which are pointed out in Fig~\ref{fig:topics_wordclouds}. The meaning of the topics or macro-topics is contextualized in \ref{sec:context}. 

\par Finally, by following the procedure described in the previous section, we construct the Agendas as distributions of topics coverage in media (Media Agenda) and public searches (Public Agenda) in Google and Twitter.

\subsection{The Media and Public Agendas}
\label{sec:Quantification}

\begin{figure}[htbp]
\includegraphics[width = \textwidth]{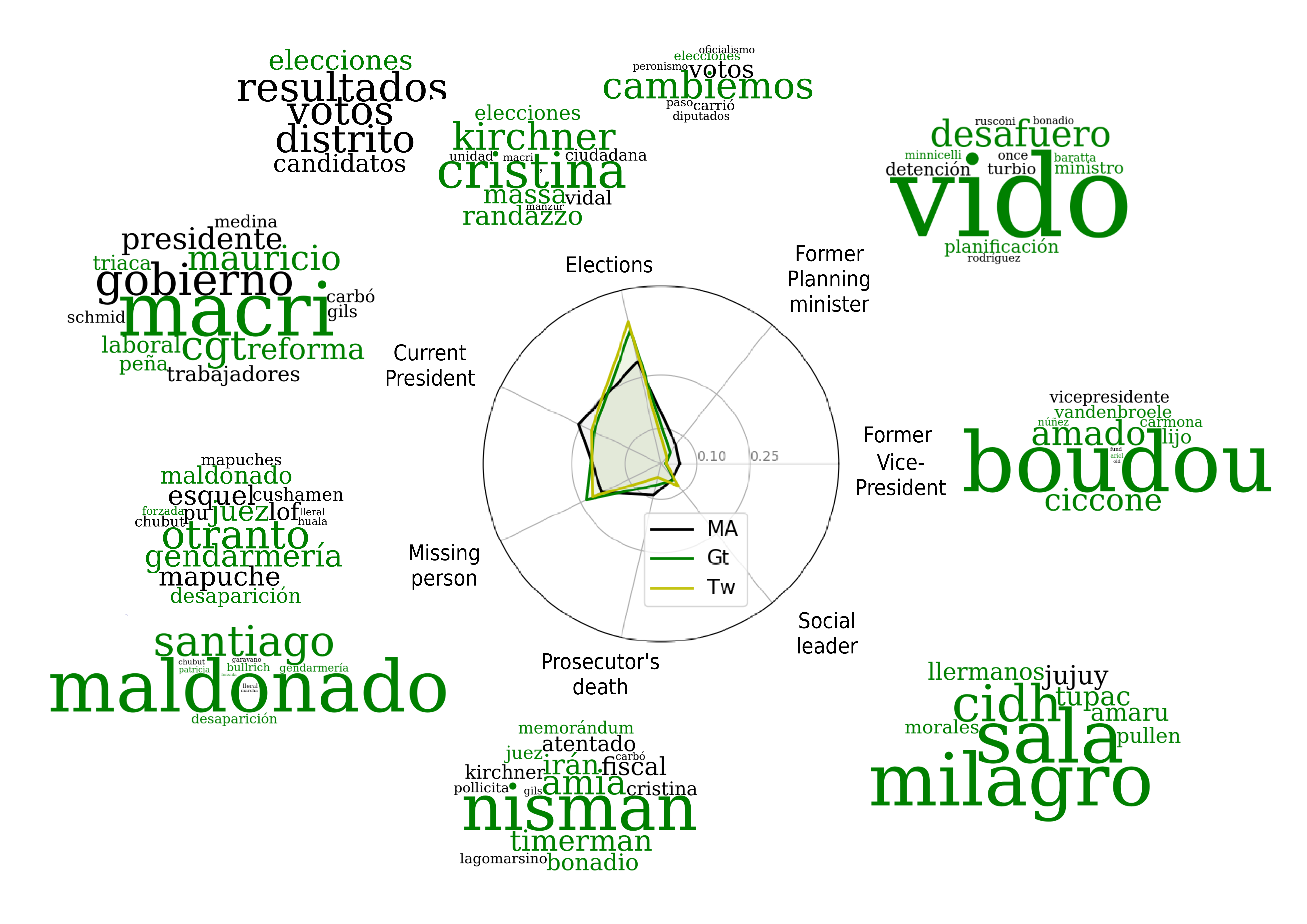}
\caption{\textbf{Radar plots of the Media and Public Agendas represented by the ten topic (then collapsed to seven) distribution and their corresponding word clouds.} The Public Agenda is represented both by Google Trends (GT) and Twitter (Tw). Topic names are introduced together with the word clouds containing the most important keywords involved in the definition of each topic. In green color we show the keywords used to define the topics in the Google Trends and Twitter queries (see Table~\ref{table:gt_queries}) and therefore in our construction of the Public Agenda.}
\label{fig:topics_wordclouds}
\end{figure}

\par In Fig~\ref{fig:topics_wordclouds} we show the Media (MA) and Public Agendas (PA) (discriminated by  Google Trends (GT) and Twitter (Tw)) in a seven topic decomposition of the whole corpus using radar plots to represent the average distribution. In this figure we also show the word clouds of the keywords that define each of the ten original topics, where the size of the word reflects its importance in the topic definition. In green color, we point out the words involved in the Google Trends and Twitter queries in order to construct the Public Agenda. The queries employed are also specified in Table~\ref{table:gt_queries}.

\begin{table}[htbp]
\centering
\begin{tabular}{l|p{7cm}}
Topic name & Google Trends query + Twitter keywords (underlined)\\ \hline
Elections & \underline{elecciones} + \underline{cambiemos} + cristina \underline{kirchner} + \underline{massa} + \underline{randazzo} \\
Missing person & santiago \underline{maldonado} + juez \underline{otranto} + patricia bullrich + \underline{gendarmería} + \underline{desaparici\'on} forzada \\
Former Planning minister & de \underline{vido} + \underline{desafuero} + ministro de planificaci\'on + \underline{minnicelli} + \underline{baratta} \\
Current President & mauricio \underline{macri} + \underline{cgt} + reforma \underline{laboral} + peña + \underline{triaca} \\
Social leader & milagro \underline{sala} + \underline{cidh} + \underline{tupac} \underline{amaru} + \underline{pullen} \underline{llermanos} + \underline{morales} \\
Prosecutor's death & \underline{nisman} + \underline{amia} + \underline{memor\'andum} con ir\'an + \underline{timerman} + juez \underline{bonadio} \\
Former Vice-President & amado \underline{boudou} + \underline{ciccone} + ariel \underline{lijo} + \underline{vandenbroele} + núñez \underline{carmona} \\
\end{tabular}

\caption{Queries used in Google Trends and words employed in Twitter search (underlined) in order to build the Public Agenda}
\label{table:gt_queries}

\end{table}

\par We can see that both GT and Tw look similar in this representation, but they show specific differences with the Media Agenda. For instance, a greater interest of the audience in the topic \emph{Missing person} than the media is observed, or inversely, a lower interest in the topic \emph{Prosecutor's death} takes place. However, this static representation is not able to show the complex dynamics of the agendas evolution and the importance of punctual and specific facts which can erase or amplify their differences. 

\par
\begin{figure}[htbp]
\includegraphics[width = \textwidth]{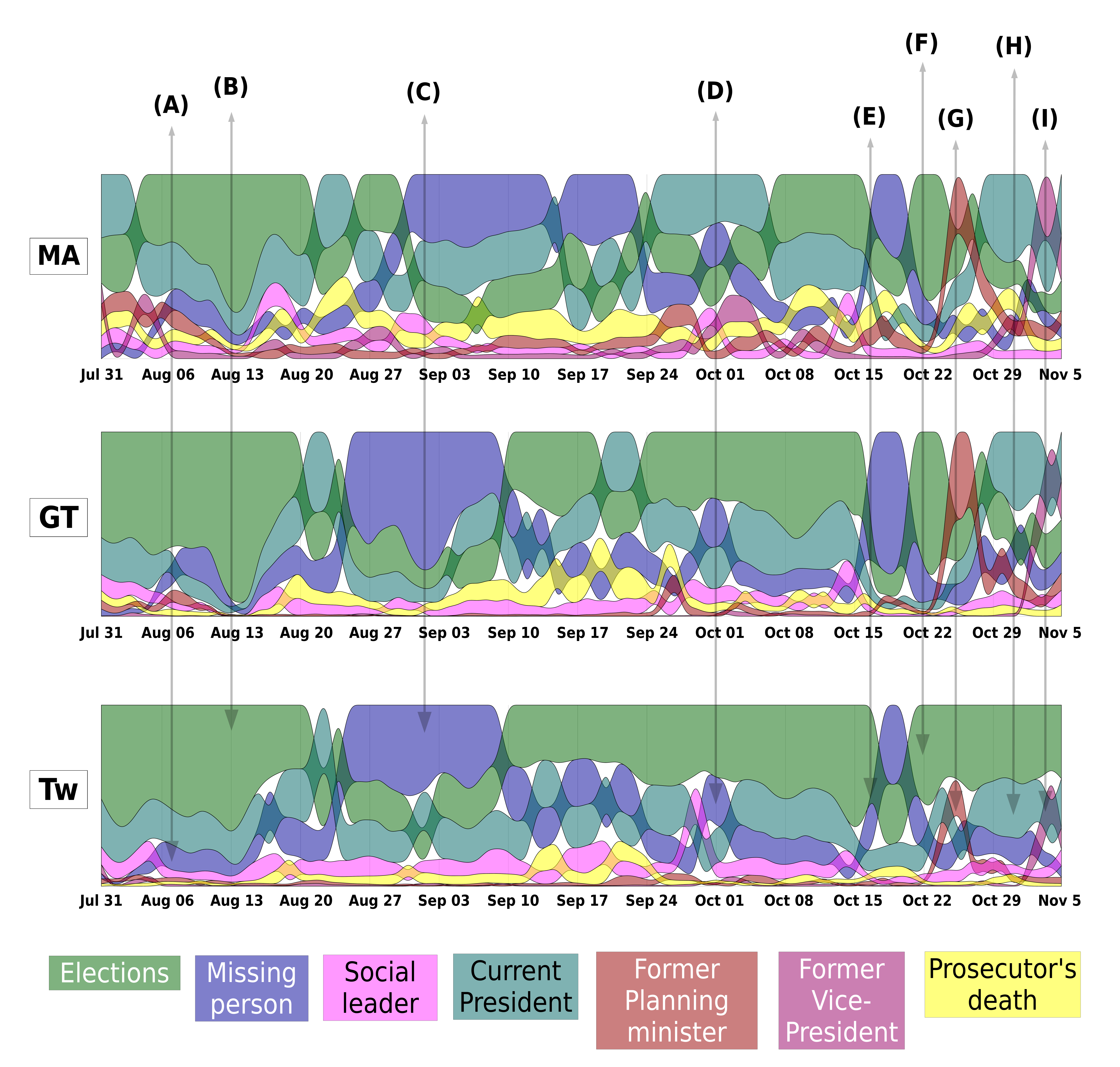}
\caption{\textbf{Bump graph of the time-dependent Media (MA) and Public Agendas extracted from Google Trends (GT) and Twitter (Tw)}. Widths and rankings of the curves encode topic's relative weight. Also, some important events related to the topics are pointed out:
\textbf{A}: First news about Santiago Maldonado's disappearance (Missing person);
\textbf{B}: Primary elections;
\textbf{C}: March one month after Santiago Maldonado's disappearance;
\textbf{D}: March two months after Santiago Maldonado's disappearance;
\textbf{E}: Appearance of Santiago Maldonado's body;
\textbf{F}: General elections;
\textbf{G}: Julio De Vido's detention;
\textbf{H}: Debates on labor reform;
\textbf{I}: Amado Boudou's detention (Former vice-president).
A more detailed explanation is given in \ref{sec:context}.}
\label{fig:all_agenda}
\end{figure}

 The time evolution of the agendas in the topic space is shown as bump charts of the Agendas in Fig~\ref{fig:all_agenda}. The bump chart provides a clear visualization of the relative weight of the topics at the same time with their ranking. In Fig~\ref{fig:all_agenda}, we also highlight some important events related to the dynamics of the topics.
It is possible to appreciate how the main topic changes with time and have a glance of the qualitative differences between the agendas. In particular, it can be seen some differences between the Public Agendas that were not observed in Fig~\ref{fig:topics_wordclouds}, as for instance, the persistence of main topics is longer in Twitter than in Google Trends. This is more evident at the end of the analyzed period, where the topics discussed in Google Trends show more response to change in Media Agenda than in Twitter, maybe due to the existence of a different pattern of interaction in the social network, to which a deeper analysis could be devoted in future works.

\par The linear correlations between the same topics in MA and PA were also calculated. In all cases, we found that the correlations are positive and statistically significantly, as it is shown in Table~\ref{table:gt_all_correlation}.
We interpret this as a validation of the topics found in the corpus and the keywords that describe it. Even though we are particularly interested in those periods where the Agendas differ, it is expected that the media and public interests should generally follow a similar a pattern, mainly driven by external events.

\subsection{Agenda diversity}
\label{sec:Diversity}

\par How dominant is a main topic? Is the degree of dominance of a given topic in the Media Agenda reflected in the Public Agenda? Diversity is a key variable when dealing with multiple issues \cite{boydstun2014importance}, due to the fact that it tells us how the attention is distributed across the different topics of discussion. As was proposed in \cite{boydstun2014importance}, we use the normalized Shannon entropy $H$ to quantify the diversity within our framework.

\par
\begin{figure}[htbp]
\centering
\includegraphics[height=0.7\textheight]{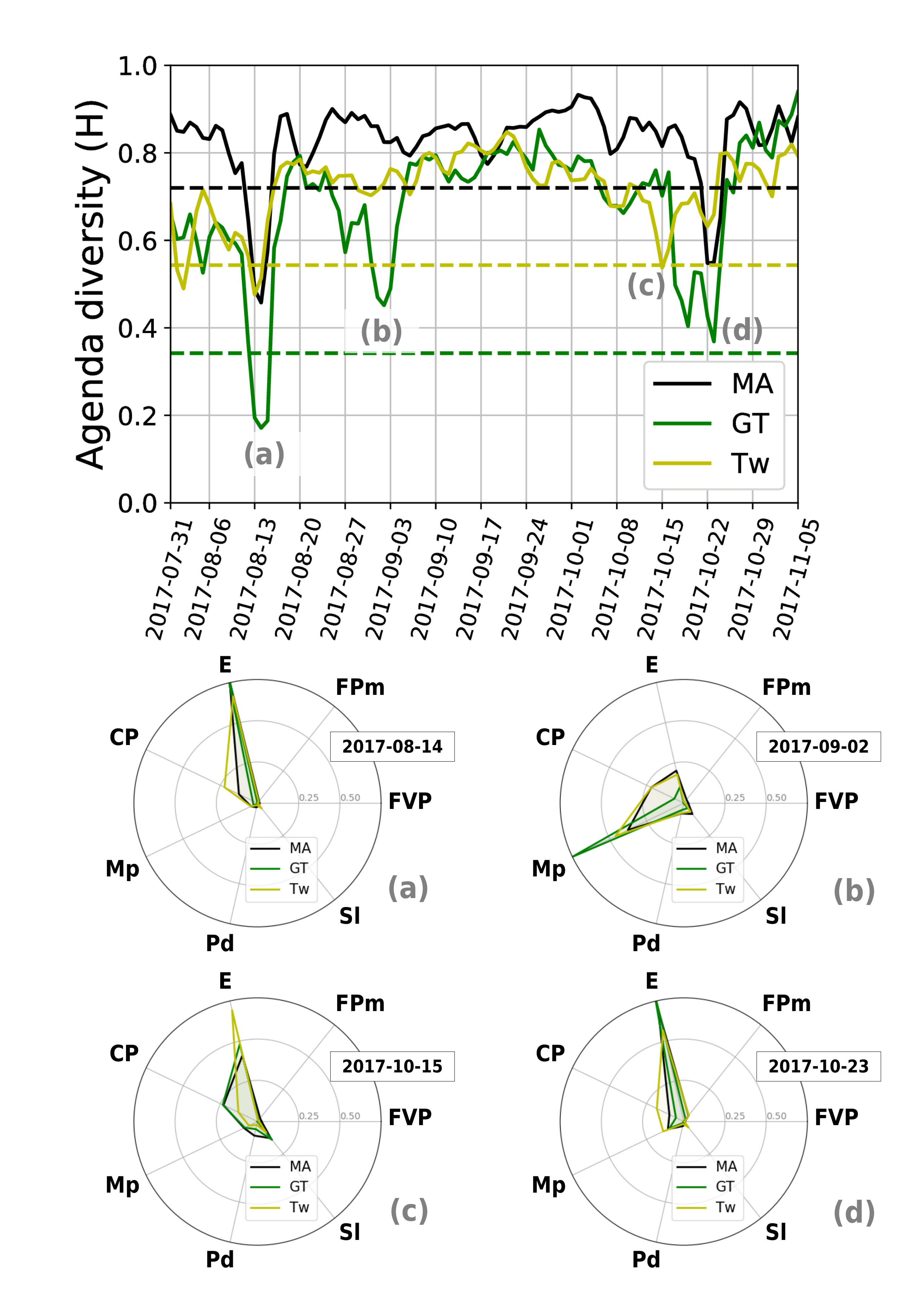}
\caption{\textbf{Shannon entropy (H) as a measure of agenda diversity.} The Public Agenda shows a less diverse behavior than the Media Agenda as can be seen in the top figure. The horizontal lines correspond to the lower inner fences of each signal in order to identify outliers. The related radar plots show the agenda at the selected days where the time series exhibit dropouts (points a-d), indicating that the most important topic catches most of the public's attention. \textbf{E}: Elections; \textbf{FPm}: Former Planning minister; \textbf{FVP}: Former Vice-President; \textbf{Sl}: Social leader; \textbf{Pd}: Prosecutor's death; \textbf{Mp}: Missing person; \textbf{CP}: Current President. }
\label{fig:shannon_entropy_agendas}
\end{figure}

 In Fig~\ref{fig:shannon_entropy_agendas} we can see the value of $H$ as a function of time for the three agendas. It is important to pay attention to those periods of time when the diversity is lower than usual. This effect is notoriously more pronounced in the Public Agenda giving by GT, and in particular in four specific days when four local minimums of the Shannon entropy can be detected. Three of them are outliers as defined in \ref{sec:measures}, two of them from GT and one from Tw. The other one has not been identified as an outlier but it is a pronounced minimum and therefore a point of interest in our description. 

\par A lower value in the agenda diversity is due to the fact that the most important topic attracts practically all the attention of the public and the media, collapsing the agenda to one of the issues involved.
In the radar plots included in Fig~\ref{fig:shannon_entropy_agendas} we can see how two of these outliers (\textbf{a} and \textbf{d}) belong to the topic \emph{Elections}. They are related to the primary and general legislative elections that took place in August $13^{th}$ and October $22^{nd}$ respectively.
In all the agendas these points were detected as outliers except point (d) in Twitter Agenda. Why is that? The radar plot of the Twitter agenda for this day displays an association between the topic \emph{Elections} and the \emph{Current President}, decreasing the importance of this topic.
Discussions in Twitter about elections appear also in point (c), when the other agendas seem to be more diverse. 
On the other hand, and despite not being classified as outlier, we also focus on point (b) because the Shannon Entropy in the Google Agenda displays a minimum (collapsing agenda) which is not corresponded neither in the Media nor in the Twitter Agendas. Crawling in the context, we see that it belongs to the topic \emph{Missing person} and this date corresponds to the rally that took place one month after the disappearance of Santiago Maldonado, as was mentioned in section \ref{sec:topic-decomposition}. 

\par From the measure of $H$ we have also observed that the median of the Public Agenda diversity is statistical significant lower than the median of the Media Agenda.
Specifically $H_{GT} = 0.73$ and $H_{Tw} = 0.74$ are statistically significantly lower than $H_{MA} = 0.85$ with $p < 10^{-18}$, while there is no significant difference between the first two. 
However, from Fig~\ref{fig:shannon_entropy_agendas} we can see that GT shows more abrupt dropouts in the diversity in response to specific events.
From all this analysis we can conclude that given a finite set of topics, the Public Agenda is less diverse than the Media Agenda, because the public seems to focus on the most important topics than the media can do, maybe due to editorial decisions.

\subsection{Distance between Media and Public Agendas}
\label{sec:Distance}

 \par
\begin{figure}[htbp]
\centering
\includegraphics[height = 0.7\textheight]{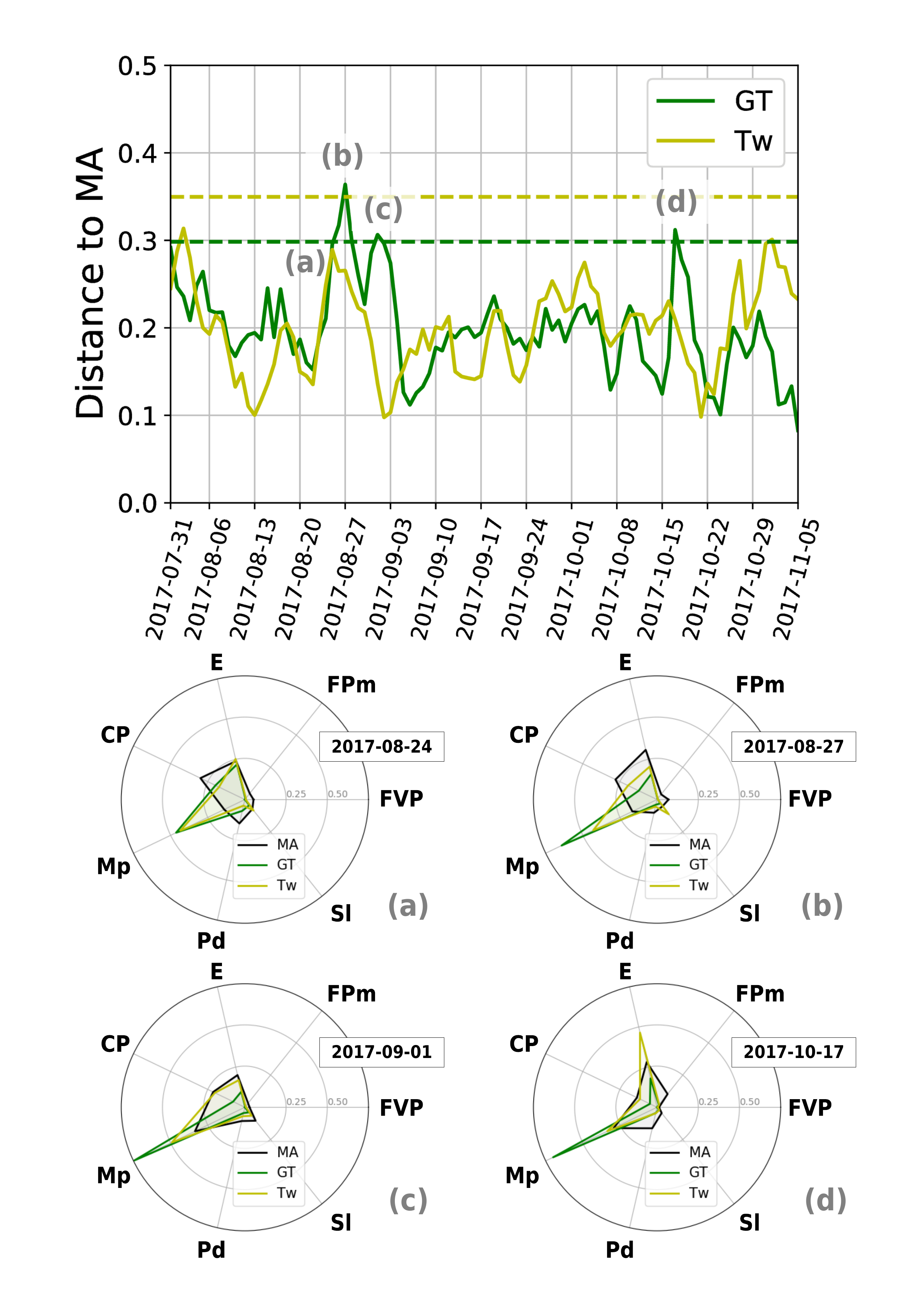}
\caption{\textbf{Jensen-Shannon distance between the Media and Public Agendas as a function of time} (with upper inner fences pointed out). Larger distances are due to a greater interest of the audience in the topic \emph{Missing person} which decreases the interest in other topics. On the other side, the Media Agenda still keeps certain degree of diversity. \textbf{E}: Elections; \textbf{FPm}: Former Planning minister; \textbf{FVP}: Former Vice-President; \textbf{Sl}: Social leader; \textbf{Pd}: Prosecutor's death; \textbf{Mp}: Missing person; \textbf{CP}: Current President.}
\label{fig:jensen_shannon_gt}
\end{figure}

 While the diversity is a property of each distribution, a natural question  when comparing different distributions is how similar they are. Given our descriptions of the agendas as time-evolving distributions, we can compare them by computing the Jensen-Shannon distance (see \ref{sec:measures} for details). In this context, outliers in selected dates will correspond to divergences between the Media and Public Agenda: Specific events when the public interests do not match with media offer.
In Fig~\ref{fig:jensen_shannon_gt} we show the Jensen-Shannon distance between Media and Public Agendas as a function of time. We focus on three points that seem to be relevant enough. In all cases, the topic distributions at these days displayed show that the increment in the distance between agendas is due to a greater interest of public opinion in the topic \emph{Missing person}. 
\par Points \textbf{(c)} and \textbf{(d)} of Fig~\ref{fig:jensen_shannon_gt} show that both the public and the media highlight this topic, but the media do not disregard other topics, so the corresponding distance between them can be interpreted as lack of diversity in Public Agenda as discussed in the last section.
On the other hand, points \textbf{(a)} (we take this point due to be a local maximum despite not being an outlier) and \textbf{(b)} show a major interest of the public in the topic \emph{Missing person} which it is not reflected in the Media. 
In Fig~\ref{fig:all_agenda} we can see that this topic becomes the most important in public interests (both in GT and Tw) days before that it happens in the Media Agenda. This fact can be associated with a social networks (like Facebook and Twitter) campaign in favor of the appearance of Santiago Maldonado (``The missing person") that took place on August $26^{th}$. This campaign was massive and initially underestimated by the main media outlets in Argentina. 

\par Finally, it is important to say that the Jensen-Shannon distance, together with the measurement of agenda diversity given by the Shannon entropy, gives an insight of independent behavior of the audience and the media in certain particular dates. Its identification can be a starting point to study the media reaction to a change in audience interests. 

\subsection{Agenda bias in different media outlets}
\label{sec:Agenda-bias}

\par
\begin{figure}[htbp]
\centering
\includegraphics[scale = 0.13]{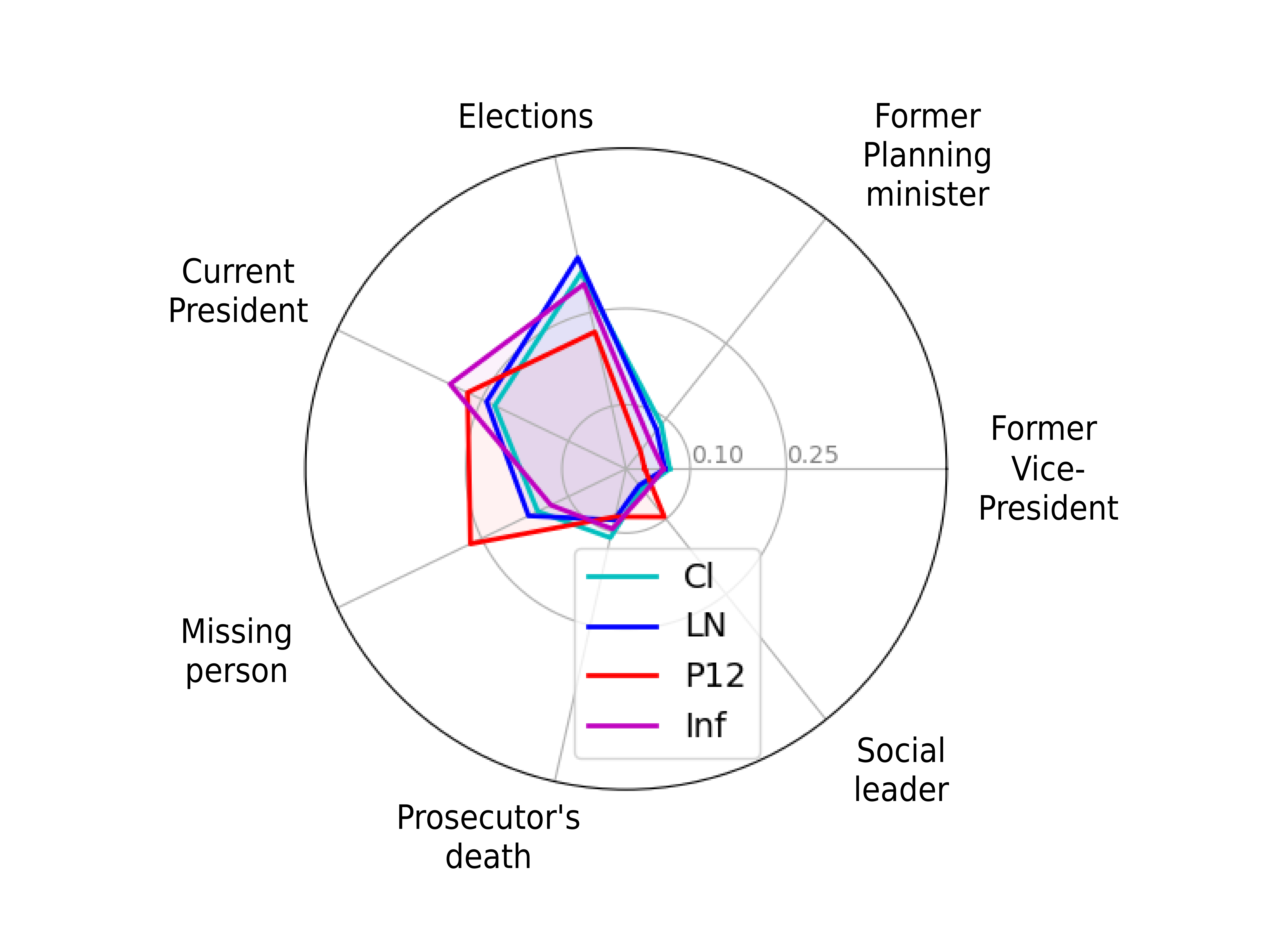}
\caption{\textbf{Radar plot of the average distributions.} The figure shows in a qualitative way the bias in the different newspaper agendas. For instance, the greater interest of P\'agina 12 (P12) in the topic \emph{Missing person} and its slightly lower coverage in the \emph{Former Planning minister} respect to the other newspapers.}
\label{fig:news_agenda}
\end{figure}

 In this section we leave aside the Public Agenda as an unified corpus and we study the composition of the Media Agenda on each media outlet. In Fig~\ref{fig:news_agenda} we show the agendas of the newspapers represented as radar plots analogously to the one shown in Fig~\ref{fig:topics_wordclouds}. It is important to recall that the topics are the same that were introduced in the word clouds of Fig~\ref{fig:topics_wordclouds}, but at the time of computing the topic weights, the articles were discriminated by newspaper. 

\par Fig~\ref{fig:news_agenda} shows qualitatively the differences between the newspaper agendas.
For instance, we can see how the newspaper called P\'agina 12 gave more importance to the topics \emph{Missing person} and \emph{Social leader}, while it reduces to minimum the coverage of the topic \emph{Former Planning minister} as the others did.
Difference in the coverage is known as \emph{coverage bias} \cite{dallmann2015media}.

\par The observed bias is consistent with was expected from the analyzed newspapers and reflects the highly polarized political climate observed in Argentinian society. During the administration of Cristina Fern\'andez de Kirchner (2007-2015), the government confront with several news organizations. It led to media outlets such as Clar\'in, La Naci\'on and the news portal Infobae to be very critical of the Fern\'andez's administration, emphasizing the allegations of corruption related to it, as can be seen in the importance given to the topics \emph{Former Planning minister} and \emph{Former Vice-President}.
On the other hand, P\'agina 12 has an opposite ideological bias \cite{zunino2010cobertura, zunino2016assessment}, supporting the former administration of Cristina Kirchner and therefore being very critical with the current Mauricio Macri's administration, doing special emphasis on issues related to human rights, as can be again observed in the coverage given to the topics \emph{Social leader} and \emph{Missing person}. 
\par Given the importance of the topic \emph{Missing person}, we will devote the next section to the analysis of it.

\subsection{A single topic study: Coverage bias and complex feedback dynamics}
\label{sec:single-topic}

\par In previous section we have seen that a single topic was responsible for agenda's diversity reduction, temporal dissimilarity between public and media agenda and in coverage bias between different newspapers: The topic \emph{Missing person}. Besides, this is the most adequate one to be discussed because:
\begin{itemize} 
\item It caused a great impact in both the media and the audience;
\item its coverage fully deploys along the time lapse analyzed in this manuscript (see \ref{sec:context}).
\end{itemize}

\par
\begin{figure}[htbp]
\includegraphics[width = \textwidth]{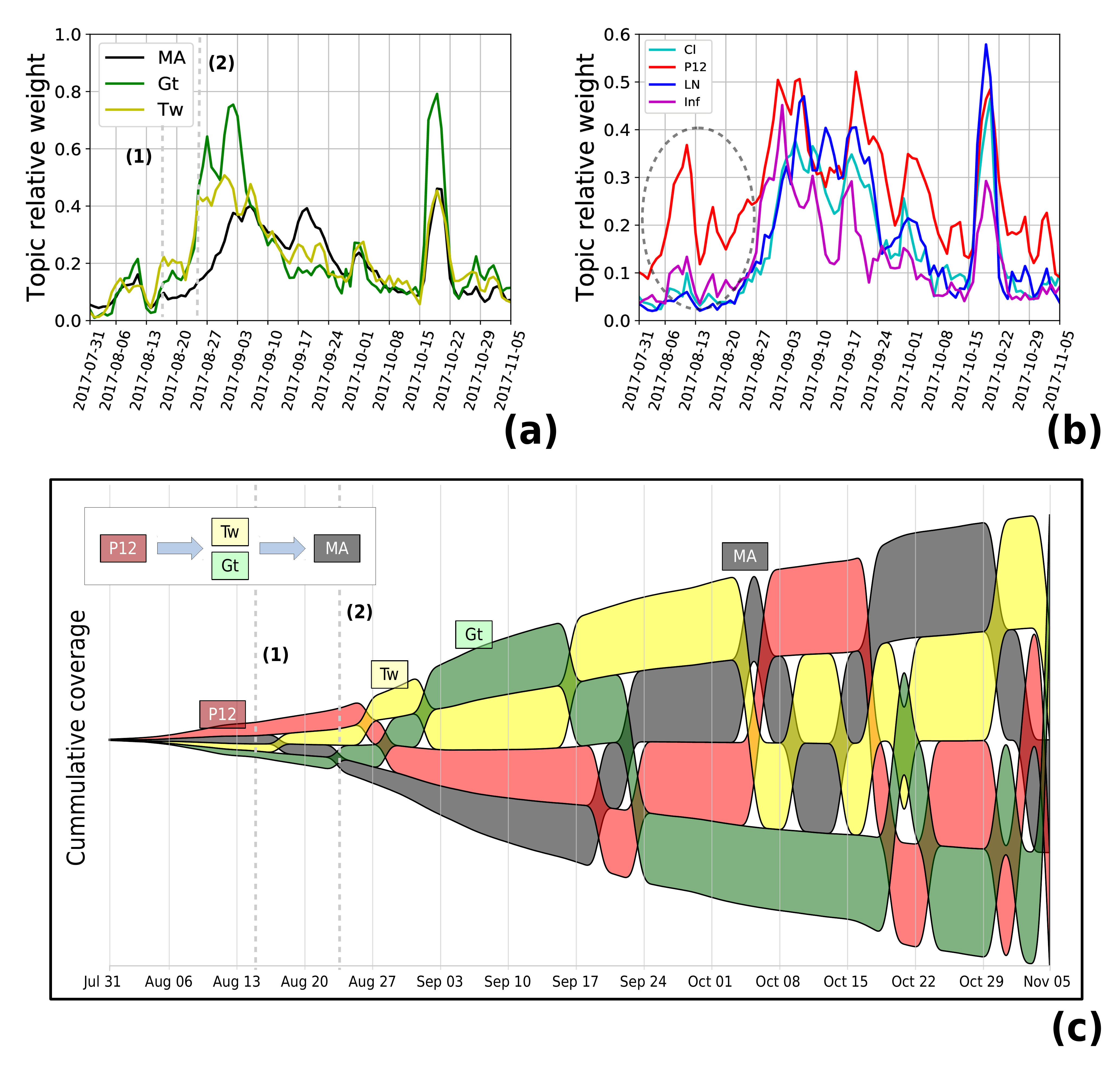}
\caption{\textbf{Non-trivial agenda interplay in the topic Missing person.}
The temporal profiles of panel (a) show that the Public and Media Agendas seem to differentiate around August $15^{th}$ (vertical grey line (1)) and the audience increases abruptly its interest in the topic around August $24^{th}$ (grey line (2)).
However, when we see the media outlet separately at panel (b), we observe that P\'agina 12 displays a greater coverage at the beginning of this topic (pointed out with a dashed circle), before the increasing of the public attention. 
By calculating the cumulative coverage, represented as a bump chart in panel (c), we suggest that the topic was first set by P\'agina 12 and then the audience interest seems to cause the coverage of the rest of the media.
}
\label{fig:Maldonado_setagenda}
\end{figure}

 In Fig~\ref{fig:Maldonado_setagenda} panel (a), we show the relative weight of the topic \emph{Missing Person} for both the Media and Public Agendas. 
After the initial coverage, the agendas seem to differentiate around August $15^{th}$, when the topic starts to become more important in the Public Agendas than in the Media one. 
Around August $24^{th}$, the topic abruptly increases in the audience interests while the reaction in the media is slower. This date is very close to August $26^{th}$, when a campaign in social media took place. After that event, the media increases its coverage about the topic. This behavior led us to the following question: Is this a case where the social media impose the media agenda? 

\par When we look closer at the media coverage we can see that not all the media outlets had the same behaviour. Fig~\ref{fig:Maldonado_setagenda} panel (b) shows the temporal profile of the topic \emph{Missing person} for each newspaper.
Notoriously, P\'agina 12 was the newspaper which rapidly paid attention to this topic since its beginning, pointed out with a dashed circle in panel (b). The range of dates where this greater coverage takes place anticipate the interest of the public shown in panel (a).

\par In order to integrate the information of the panel (a) and (b) of Fig~\ref{fig:Maldonado_setagenda}, we calculate the cumulative coverage of the topic \emph{Missing Person}, which is displayed in Fig~\ref{fig:Maldonado_setagenda} panel (c).
We define the cumulative coverage of the topic \emph{Missing Person} as the numerical integration between the initial date and the current date (normalized by the total coverage) of the topic temporal profile in the Media, Google and Twitter agenda, in addition to the agenda of P\'agina 12.
This quantity shows us how the media and public attention have been accumulated since the first events.

\par Fig~\ref{fig:Maldonado_setagenda} panel (c) suggests a complex agenda-setting dynamics within the topic \emph{Missing Person}: The least sold newspaper (P\'agina 12) triggered a public debate via a positive feedback mechanism reinforced by reiterative Google searches and discussions in social networks. Due to this increasing public interest, the rest of the media were forced to pay attention to this subject and finally, the topic becomes also prominent to the Media Agenda. 

\par Beside the analysis performed above, there are two important facts that must be mentioned about the topic \emph{Missing Person}: First, the disappearance of a person is a very sensitive issue in the Argentinian society (due to the memories of the missing persons cases during the last dictatorship), which can explain why this particular topic triggered the audience interest; and second, as was previously mentioned, P\'agina 12 was particularly interested in covering this topic since the beginning (due to its opposition to the current president administration), while the rest of the media did not until the topic was prominent to the Public Agenda.

\section{Conclusion}
\label{sec:Conclusions}
 
\par The mass media plays a fundamental role in opinion formation and therefore, it is of vital importance to have an accurate quantitative description of the Media and Public Agenda and their relationship in the framework of agenda-setting theory. In this work we develop a framework, through the implementation of a topic detection algorithm, to describe Media and Public Agendas as distribution of weights (which measures the importance of each topic) which evolves in time in a common topic space which emerges intrinsically from the corpus. Within this framework, it is possible to develop proper metrics in order to measure properties such as diversity or distance between agendas.

\par Specifically, we have found that a very attractive topic focuses the attraction of the public more than observed in media, which keeps a certain degree of diversity and a wider range of topics' interest. Interestingly, we also show that the distances between agendas can be employed to rapidly detect those periods when the public may have an independent behaviour respect to the media.

\par The same methodology allow us to compare the agendas of analyzed newspapers and quantify the coverage bias in the different topics. In particular, we detect important differences in the topic \emph{Missing person}, where the respective bias of the newspapers against and on favour of the actual government's administration seems to be relevant to explain them.

\par Also, the analysis of the topic \emph{Missing person} presented in Fig~\ref{fig:Maldonado_setagenda} shows concrete evidence of a complex dynamics agenda-setting case, in which not all the newspapers play the same role in Media Agenda and it is produced a feedback dynamics between newspapers and social networks which should be analyzed carefully within the framework of agenda-setting theory.

\par Finally, we consider that this work could be used as an starting point in developing data-driven mathematical models about the interaction between mass media and society, given the traditional approaches coming from statistical physics \cite{crokidakis2012effects,gonzalez2012model, moussaid2013opinion, rodriguez2010effects, pinto2016setting}, where much of the models lack in being contrasted with real data. Future works may include a more systematic study and its extension to international media, a deeper study combining topic detection and sentiment analysis, and a more quantitative analysis about causality.


\newpage
\appendix

\section{The Media and Public Agenda construction}
\label{sec:agenda_construction}

\subsection{The Media Agenda}

\par The news articles were described as numerical vectors through the \emph{term frequency - inverse document frequency (tf-idf)} representation \cite{xu2003document}. 
Given the set of terms contained in the corpus, after removing non-informative words such as prepositions and conjunctions, the tf-idf algorithm represents the \textit{i}-document as a vector $v_i = [x_{i1}, x_{i2}, ... , x_{it}]$, where the component $x_{ij}$ is computed by the Eq.~\ref{ec:tfidf}, $\textrm{tf}_{ij}$ is the number of times the \textit{j}-term appears in the \textit{i}-document, $d$ is the number of documents in the corpus, and $n_j$ is the number of documents where the \textit{j}-term appears. 
Each vector is then normalized to unit Euclidean length. 
Once the document vectors are constructed, we put them together in a document-term matrix (\emph{M}), which has dimensions of number of documents in the corpus ($d$) by number of terms ($t$).

\begin{equation}
x_{ij} = \textrm{tf}_{ij} \cdot \textrm{idf}_{j} = \textrm{tf}_{ij} \cdot [1 + \textrm{log}(\frac{1 + d}{1 + n_j})] 
\label{ec:tfidf}
\end{equation}

\par In order to detect the main topics in the corpus, we perform \emph{non-negative matrix factorization (NMF)} \cite{xu2003document, lee1999learning} on the document-term matrix (\emph{M}). A topic is defined as a group of similar articles which roughly talks about the same subject. 
NMF is an unsupervised topic model which factorizes the matrix \emph{M} into two matrices \emph{W} and \emph{H} with the property that all three matrices have no negative elements (see Eq.~\ref{eq:nmf}).
This non-negativity makes the resulting matrices easier to inspect, and very suitable for topic detection. 
\begin{equation}
M^{(d \times t)} \sim H^{(d \times k)} \cdot W^{(k \times t)}
\label{eq:nmf}
\end{equation}
Such as the resulting matrix \emph{H} has dimensions of number of documents by $k$ and matrix \emph{W} has dimensions of $k$ per number of terms, the number $k$ is therefore interpreted as the number of topics in the documents and it is a parameter that must be set before the factorization.
In this work, we arbitrarily set $k=10$, based on our knowledge of the corpus.
Since the factorization of Eq.~\ref{eq:nmf} usually can not be made exactly, it is approximated by minimizing the reconstruction error, i.e. the distance between matrix \emph{M} and its approximated form $\tilde{M} = H \cdot W$. The NMF factorization was made through the python module \emph{scikit-learn} \cite{scikit-learn}.

\par The matrix \emph{H} is the representation of the documents in the topic space. We normalized its rows to unit $l_1$-norm in order to view their components as a degree of membership of a given document in the set of topics. In particular, the index of the largest component tells us which is the most representative topic of the document.
On the other hand, \emph{W} gives the topics representation in the original term space. The largest components of a row give the most representative words of each topic, which we call keywords, and therefore an insight of what the topic is talking about.

\subsection{The Public Agenda}

\subsubsection{Google Trends tool}

\par We describe here the procedure in order to get information from Google Trends, employed in the construction of the Public Agenda. As was mentioned in the main text, we employed the keywords provided by the NMF algorithm in order to construct the respective queries. The Google Trends tool provides the relative amount of Google searches, normalized by the maximum value observed in the query. The relative searches are re-scaled to 0 to 100, so the query with largest searches is assigned the value of 100. In spite of the fact that we don’t have access to the absolute value of searches, we can obtain the relative ones, which is our main interest.

\par It should be mentioned that Google Trends allows only to compare five topics at the same time, so in order to obtain the information of the 7 topics involved in the analysis, we repeat the query on the same range of dates by keeping the most important topic of the period in the query (which was always easily identified) and replacing the rest of the topics with the other two missing ones. 
The queries were performed on a three-day sliding window, in an analogous way when we employed a three-day linear filter when we analyzed the time series of the newspaper topics as it as mentioned in the main text.

\par For each query performed, we obtain the mean value of searches relative to the most important topic on the period and we assign this values to the middle date. With the relative values of the 7 topics, after a proper normalization, we were able to construct the time evolving distribution date by date, which we finally called as the Public Agenda.
The average of the time-evolving agenda was used to construct the radar plot of Fig.~\ref{fig:topics_wordclouds}. In table \ref{table:output_queries}, we provide the average outputs of the Google Trend queries for each topic (which were used to construct this last figure).

\subsubsection{Twitter}

\par In the case of Twitter, its advanced search tool allowed us to use as input all the keywords of the 7 topics at the same time at a particular date. In consequence, by taking a sample of tweets, we obtain a relative amount of tweets devoted to the topics involved in our work. 
At a particular date, we observe that the distribution of tweets stabilizes in a sample of about 200 tweets. So we take at least 250 tweets per day, and use this information in order to estimate the distribution of the topics in Twitter (i.e. the Public Agenda in this social media). During all the studied period we collected a total of 24360 tweets as was mentioned in the main text. The distribution of tweets over the 7 topics is pointed out in table \ref{table:output_queries}. This information was used to construct the radar plot of Fig.~\ref{fig:topics_wordclouds}.

\begin{table}[htbp]
\centering
\begin{tabular}{lcccccccc}
& E & Mp & CP & FPm & FVP & Sl & Pd\\ \hline
Google Trends & 0.386 & 0.233 & 0.209 & 0.041 & 0.013 & 0.057 & 0.061\\
Twitter & 9987 & 5218 & 5300 & 612 & 352 & 1940 & 951
\end{tabular}
\caption{Example of output of Google Trends and amount of tweets per topic.}
\label{table:output_queries}
\end{table}

\subsection{Correlation between topics temporal profiles}

In Table~\ref{table:gt_all_correlation} we show the linear correlation between the topic temporal profiles from the Public Agenda and their counterparts in the Media Agenda.

\begin{table}[h!]

\centering
\begin{tabular}{lccc}
Topic name & Correlation MA and GT & MA and Tw & GT and Tw \\ \hline
Elections & \textbf{0.81} & \textbf{0.59} & \textbf{0.75} \\
Missing person & \textbf{0.68} & \textbf{0.76} & \textbf{0.89} \\
Former Planning minister & \textbf{0.92} & \textbf{0.82} & \textbf{0.87} \\
Current President & \textbf{0.77} & \textbf{0.75} & \textbf{0.63} \\
Social leader & \textbf{0.49} & \textbf{0.25*} & \textbf{0.57} \\
Prosecutor's death & \textbf{0.56} & \textbf{0.59} & \textbf{0.75} \\
Former Vice-President & \textbf{0.90} & \textbf{0.92} & \textbf{0.97}\\
\end{tabular}

\caption{Correlation between the topic temporal profiles of the Public Agenda and their counterpart in Media Agenda.
All correlation values are statistically significant ($p < 10^{-9}$), except (*) which is significant with $p < 0.05$.}
\label{table:gt_all_correlation}

\end{table}

\section{Topic context}
\label{sec:context}

In order to provide some insights in the context of the topics analyzed to non-Argentinian readers, we provide the following information (note that some topic keywords are underlined):

\begin{itemize}

    \item \underline{Elections}: Two legislative elections were celebrated during the period in great part of Argentina: Primary elections on August $13^{th}$ and the general elections on October $22^{nd}$, 2017. A special focus was put on the elections in the Buenos Aires province, where the former President \underline{Cristina} \underline{Kirchner} participated as a senator candidate representing the alliance Unidad Ciudadana, confronting \underline{Cambiemos}, which is the alliance of the current President Mauricio Macri and the current governor of Buenos Aires province Maria Eugenia Vidal. On the other hand, two other candidates that also participated in the election were Sergio \underline{Massa} and Florencio \underline{Randazzo}.
    
    \item \underline{Current President}: Mauricio \underline{Macri} is the current Argentinian President since December 2015 and this topic is mainly composed of articles related to his administration, specially after the general elections of October $22^{nd}$, 2017, when a labour reform promoted by the government was being discussed.
    
    \item \underline{Missing Person}: \underline{Santiago Maldonado} disappeared on August $1^{st}$, 2017 after a minor clash between the \underline{Gendarmerie} (Border Guards) and a group of \underline{Mapuches} (Patagonian native population). Since that event, the Mauricio Macri's administration was accused by several people as the responsible for a forced disappearance. A very massive campaign in social media took place on August $26^{th}$, 2017 under the motto ``Where is Santiago Maldonado?", followed by two massive protest marches to the Plaza de Mayo that took place on September $1^{st}$ and October $1^{st}$. The first one had a great repercussion due to several incidents that took place during the march. The body of Santiago Maldonado was found dead on October $17^{th}$, 2017 in the Chubut river, near the place where he was seen for the last time, and the autopsy report told that Santiago Maldonado had died from ``asphyxia after being submerged", with no injuries on his body. However, due to the fact that this topic is a very sensitive one for the Argentinian society, the causes of the Maldonado's death are still being investigated.
    
    \item \underline{Former Planning minister}: Julio \underline{de Vido} was the Planning minister during the administration of N\'estor Kirchner and Cristina Fern\'andez de Kirchner (2003-2015). In 2015, he was elected to integrate the Chamber of Deputies, which finally voted to strip De Vido of his congressional immunity over corruption allegations and was immediately jailed on October $27^{th}$, 2017.
    
    \item \underline{Former Vice-President}: Amado \underline{Boudou} was the \underline{Vice-President} of the Cristina Kirchner's administration. Boudou was arrested on November $3^{rd}$, 2017 on charges including money-laundering and hiding undeclared assets.
    
    \item \underline{Social leader}: \underline{Milagro Sala} is an indigenous leader. She has been incarcerated under pre-trial detention ever since she was first detained in January 2016. She faces allegations of embezzlement related to government funding for housing projects managed by \underline{T\'upac Amaru}, her social organization. Sala accused the government of ``violating her human rights", and several people think that she is a political prisoner.
    
    \item \underline{Prosecutor's death}: Alberto \underline{Nisman} was a special prosecutor who were investigating the 1994 terror attack on the Argentine Israeli Mutual Association (AMIA), until his suspicious death in January 2015. During the period analyzed in this work, a team of experts led by the Gendarmerie (Border Guard) concluded that late prosecutor's death may have been a case of murder, not suicide.

\end{itemize}

\section{Measures}
\label{sec:measures}

\subsection{Normalized Shannon Entropy ($H$)}

\par The normalized Shannon entropy $H[p]$ referred in Eq.~\ref{eq:shannon_entropy} gives us a measure of how spread is a discrete distribution, taking the maximum value of $1$ when all outcomes are equally probable (as in the case of having a diverse agenda), and $0$ when there is just one possible outcome (when one topic of discussion dominates the agenda). 

\begin{equation}
H[p] = \frac{- \sum_{i = 1}^{N} p(x_i) * ln(p(x_i))}{ln(N)}
\label{eq:shannon_entropy}
\end{equation}

\subsection{Jensen-Shannon distance}

\par We measure the similarity between distributions via the Jensen-Shannon distance ($JSD$). 
When the similarity of the distributions is low, the distance between them is high. 
The Jensen-Shannon distance ($JSD$) is a metric between distributions based on the Jensen-Shannon divergence ($JS_{Div}$) \cite{fuglede2004jensen}, which is in turn a symmetric version of the well-known Kullback-Leibler divergence $D_{KL}$ (Eq.~\ref{eq:kl}). 
We recall that the $JSD$ has the advantage of being symmetric and also a well-defined distance, which makes it conceptually easier to deal with.
As can be seen in Eq.~\ref{eq:jensen_shannon_distance}, the $JSD$ between the distributions $P$ and $Q$ is simply the square root of the Jensen-Shannon divergence, where $M = \frac{P + Q}{2}$.

\begin{equation}
D_{KL}(P||Q) = -\sum{P(i) log(\frac{Q(i)}{P(i)})}
\label{eq:kl}
\end{equation}

\begin{equation}
JSD(P,Q) = \sqrt{JS_{Div}(P,Q)} = \sqrt{\frac{1}{2}[D_{KL}(P||M) + D_{KL}(Q||M)]} 
\label{eq:jensen_shannon_distance}
\end{equation}

\subsection{Outliers identification}

\par In order to detect these outliers we follow the popular box-plot construction proposed by Tukey \cite{tukey1977exploratory}, which is a simple data-driven method and has the advantage of making no prior assumption about the distribution of the data. 
However, it is important to remark that the constants involved in the outliers definition (see Eq.~\ref{eq:fences}) is taken from applying this method on a normal distribution.
\par In the box-plot construction a quartile division of the $N$ observations is proposed. We name $Q1$ as the lower quartile, $Q2$ the median of the distribution, and $Q3$ the upper quartile. Recall that $Q1$ ($Q3$) is defined to be the division where the 25th (75th) percent of the observations lies below (by definition, the median $Q2$ separates the distribution in two equal parts). On the other hand, the inter-quartile range $IQ$ is defined to be $IQ = Q3 - Q1$. This is the range where the bulk of the data lies inside.
We are not interesting in the visualization of the box-plot in its own but instead in its procedure to identify outliers. Therefore, from the identification of the quartiles, new quantities called fences are defined in Eq.~\ref{eq:fences}: The lower inner fence ($LIF$), the upper inner fence ($UIF$), the lower outer fence ($LOF$), and the upper outer fence ($UOF$). The fences can be interpreted as the limits of the distribution.
\begin{eqnarray}\label{eq:fences}
LIF & = Q1 - 1.5 IQ \nonumber \\
UIF & = Q3 + 1.5 IQ \nonumber\\
LOF & = Q1 - 3 IQ \nonumber \\
UOF & = Q3 + 3 IQ
\end{eqnarray}
\par We then have all the ingredients to label a point as an outlier: A point which lies above the upper inner fence is considered a mild outlier, while a point that lies above the upper outer fence is considered an extreme outlier. The same holds for the lower fences, i.e. if a point lies below the lower inner (outer) fences is considered as a mild (extreme) outlier \cite{natrella2010nist}. 
We will indicate the proper fences in each figure either when the diversity or the distance is being analyzed. We will pay attention not only to those values labeled as outliers, but also to those that are next to any of the fences despite not being strictly defined as that.

\section{Validation of NMF}
\label{sec:validation}

\par We provide here a measure of the performance of the Non-Negative Matrix Factorization algorithm (NMF) on a labeled corpus. We apply NMF on this corpus and analyze its capacity to describe the expected structure. 
The labeled corpus is made up of 400 newspapers articles composed of 4 topics (100 articles per topic). The topics talk about the US President Donald Trump, the Argentinian football club River Plate, and two of the topics analyzed in our work: the disappearance of Santiago Maldonado (the \emph{Missing person}), and articles related to Julio De Vido, the \emph{former-planning minister}. 
\par We perform a 4-fold cross validation by training the tf-idf vectorization and NMF algorithm with 300 articles and testing on the other 100. 
We first measure the quality of the algorithm by varying the number of expected topics. It is important to recall that NMF returns both the distributions of topics over the term space (which give their interpretation) and the distributions of the articles over the topic space. 
The algorithm quality could be measured by the average normalized entropy \ref{eq:shannon_entropy_sm} (where $h_{ji}$ is the degree of membership of article $j$ on topic $i$, $k$ is the number of topics, and the brackets denote average over the testing articles) of the article distributions over the topic space.
We use this quantity in order to test if the algorithm can find well-defined topics and, at the same time, it can assign as accurate as possible one of these topics to each article. In other words, a low average entropy means that most of the articles are well defined to belong to a given topic (defined by the training set).

\begin{equation}
<H_j> = <\frac{- \sum_{i = 1}^{k} h_{ji} * ln(h_{ji})}{ln(k)}>
\label{eq:shannon_entropy_sm}
\end{equation}

\par
\begin{figure}[htbp]
\centering
\includegraphics[scale = 0.35]{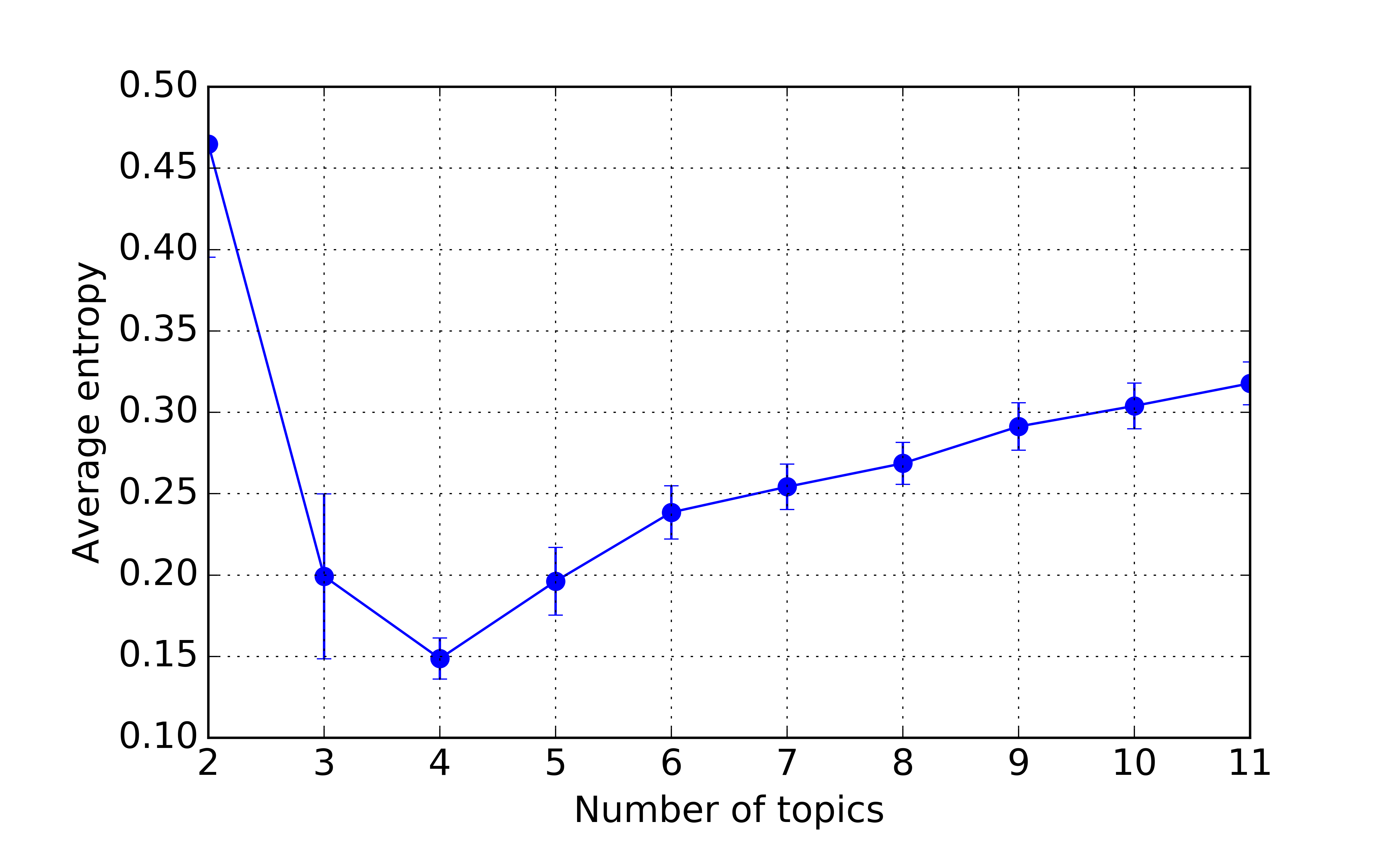}
\includegraphics[scale = 0.35]{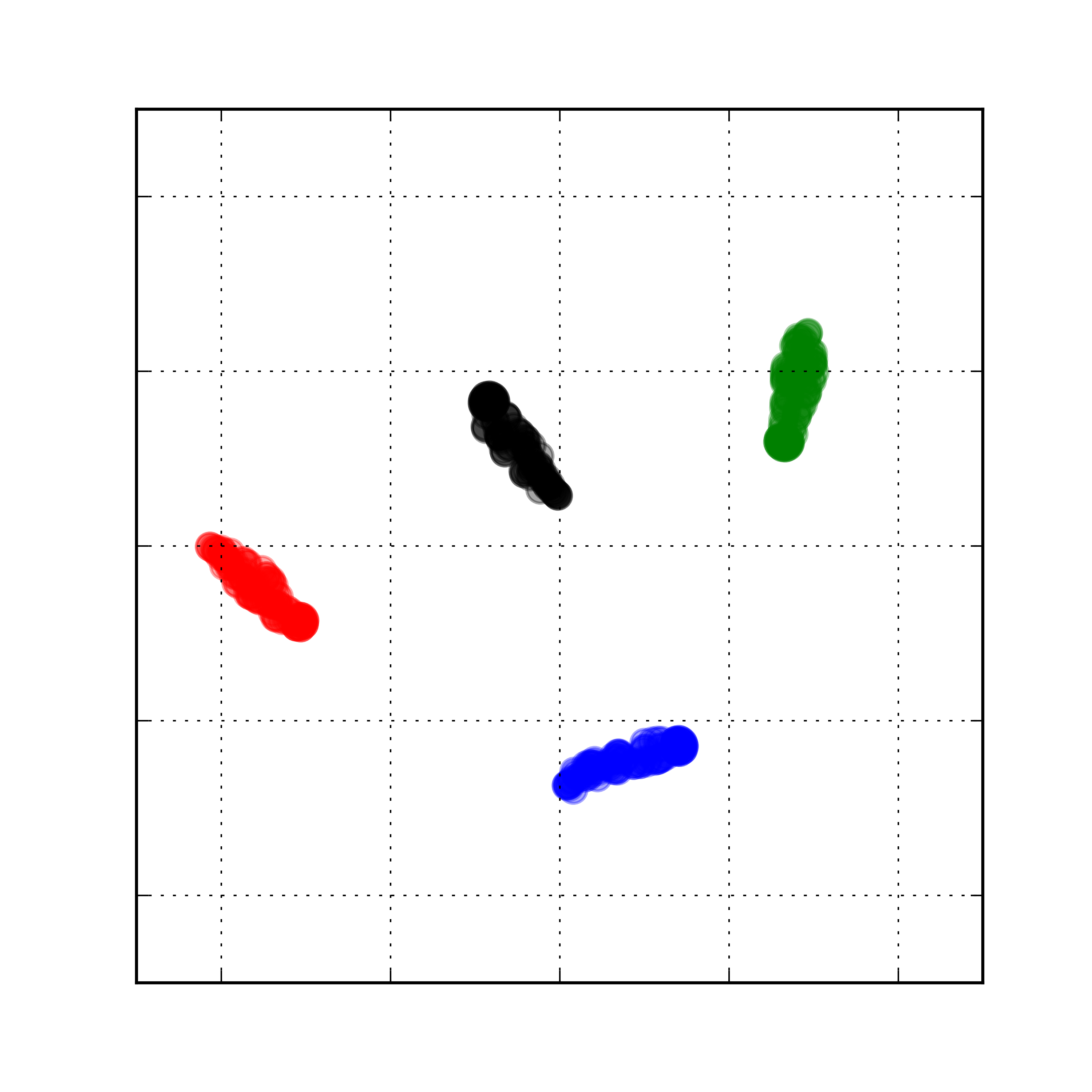}
\caption{Average entropy as function of the expected number of topics on a labeled corpus (left). t-SNE visualization of the 4-topics NMF decomposition (right). In this figure, the colors point out the original labels.} 
\label{fig:nmf_validation}
\end{figure}

 In Fig.~\ref{fig:nmf_validation}, we show the average entropy as a function of the number of topics. As can be seen, the algorithm finds that the articles have the best topic assignment when the number of topics is $4$. By setting the number of expected topics equal to $4$, we obtain the following keywords:

\begin{itemize}
    \item \emph{Santiago Maldonado}: Maldonado, Santiago, desaparici\'on, gendarmer\'ia, familia.
    \item \emph{Donald Trump}: Trump, Donald, Presidente, EE UU.
    \item \emph{Julio De Vido}: De Vido, ministro, desafuero, Kirchner, c\'amara.
    \item \emph{River Plate}: River, Gallardo Marcelo (DT), Alario Lucas (player).
\end{itemize}

\par Besides the emerging keywords are meaningful,  it is important to notice that the first keyword of each topic  is immediately associated with the predefined label.
We also compare these topics with those obtained by applying Latent Dirichlet Allocation (LDA) method \cite{blei2003latent} on the same corpus. We compare NMF and LDA results due to the fact that the last one is now the most popular method in the field of topic modelling. Both NMF and LDA were performed using the python library \emph{scikit-learn} \cite{scikit-learn}. The topics obtained with LDA are similar to those obtained with NMF unless a different ordering in the keywords.

\par In order to check the accuracy of the results obtained with NMF, we label each article by its most probable topic. Doing so, we are able to calculate the mutual information between the predicted topics and the expected ones. In this corpus, the normalized mutual information is exactly 1.00. Same results were obtained with the results obtained using LDA method. 

\par The perfect association between predicted and expected labels can be visualized by the t-SNE algorithm \cite{maaten2008visualizing}, which is widely employed in the visualization of high-dimensional datasets.
In Fig~\ref{fig:nmf_validation}, we plot the 4-dimensional article distributions over the topic space in a 2-dimensional scatter plot and color each article point by its original label. We can see that the same-color points define 4 well-localized clusters.
In conclusion, NMF has the capacity to detect the four expected topics in this corpus, and then correctly describes its structure.

\section*{Availability of data and material}
The data that support the findings of this study are available from Clar\'in at https://www.clarin.com/, La Naci\'on at https://www.lanacion.com.ar/, P\'agina 12 at https://www.pagina12.com.ar/, and Infobae at https://www.infobae.com/, but restrictions apply to the availability of these data and so are not publicly available. Google Trends and Twitter data are respectively openly available at https://trends.google.com/ and https://twitter.com/.

\section*{Competing interests}
  The authors declare that they have no competing interests.
  
\section*{Author's contributions}
PB conceived the study; SP collected and analyzed most of the data; FA collected Twitter data; FA, PB, and SP interpreted the data and prepared the manuscript; PB and COD give the final approval of the version to be published.

\section*{Acknowledgments}
  We thank Dr. A. Chernomoretz, Dr. M. Otero, Dra. V. Semeshenko, and Dr. M. Trevis\'an for bringing us a critical revision of the article.

\end{document}